# Junctionless Cooper pair transistor

K. Yu. Arutyunov[1,2] and J. S. Lehtinen[3]

[1] *National Research University Higher School of Economics, Moscow Institute of Electronics and Mathematics 101000, Moscow, Russia*
[2] *P.L. Kapitza Institute for Physical Problems RAS, Moscow, 119334, Russia*
[3] *VTT Technical Research Centre of Finland Ltd., Centre for Metrology MIKES, P.O. Box 1000, FI-02044 VTT, Finland*

**Abstract:**
Quantum phase slip (QPS) is the topological singularity of the complex order parameter of a quasi-one-dimensional superconductor: momentary zeroing of the modulus and simultaneous 'slip' of the phase by $\pm 2\pi$. The QPS event(s) are the dynamic equivalent of tunneling through a conventional Josephson junction containing static in space and time weak link(s). Here we demonstrate the operation of a superconducting single electron transistor (*Cooper pair transistor*) without any tunnel junctions. Instead a pair of thin superconducting titanium wires in QPS regime were used. The current-voltage characteristics demonstrate the clear Coulomb blockade with magnitude of the Coulomb gap modulated by the gate potential. The Coulomb blockade disappears above the critical temperature, and at low temperatures can be suppressed by strong magnetic field.

**Corresponding author**: karutyunov@hse.ru

*Introduction*

When an object with capacitance *C* is charged, its energy rises by the charging energy $E_c = q^2/2C$, where *q* is the deposited charge. It has been pointed out that if the capacitance *C* is sufficiently small, then charging by a unit charge *e* might disable electron transport through such a capacitor: the energy barrier for the next electron to enter the system is so high, that it cannot pass before the first electron leaves. To realize such a regime one should satisfy certain conditions. First, the charge *q* should be well localized within the 'capacitor' being isolated from the environment. Second, to observe such a single electron phenomenon the external disturbance should be much smaller than the charging energy. To satisfy these principal requirements at realistic temperatures (e.g. ≳ 100 mK) the system should be of sub-μm dimensions [1,2]. Typically the corresponding device, single electron transistor (SET), consists of a small central electrode 'island' isolated from the measuring circuit with two tunnel junctions (Fig. 1a). The third electrode, the gate, is used to monitor the energy of the island. If the potential of the gate $eV_{gate} > e^2/2C_\Sigma$, where $C_\Sigma$ is the effective capacitance of the system taking into consideration the capacitance of the island, capacitance to the ground, cross capacitance between electrodes, etc., then the electric current can go through the system. In the opposite limit $eV_{gate} < e^2/2C_\Sigma$ the electric conductance of the system tends to zero and one observes Coulomb blockade. Utilization of SETs have found several important applications such as the standard of electric current [3]. Utilization of superconductors enables single Cooper pair (*2e*) transport, which can be used for building of a charge qubit [4].

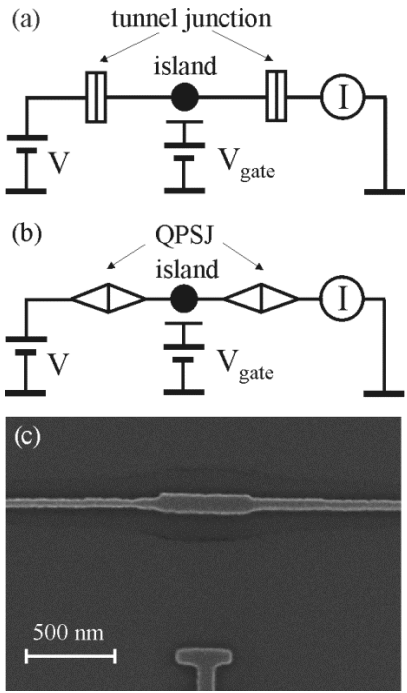

*Fig. 1.* **SET and samples**. *(a) Schematics of a conventional single electron transistor, (b) quantum phase slip transistor, and (c) scanning electron microscope image of the central part of QPSJ SET.*

It has been realized that in sufficiently narrow superconducting channels quantum fluctuations of the order parameter $\Delta=|\Delta|e^{i\varphi}$ can dramatically modify the properties of such a quasi-1D superconductor [5]. The specific manifestation of the phenomenon is called quantum phase slip (QPS). The process corresponds to momentary zeroing of the modulus $|\Delta|$ and simultaneous 'slip' of the phase $\varphi$ by $\pm 2\pi$ and leads to several non-trivial effects: finite resistance at temperatures well below the critical point [6,7], suppression of persistent currents in narrow nanorings [8], and coherent superposition of QPSs [9,10] . In particular it has been shown that superconducting titanium is the material where QPS effects do exist [10, 11, 12,13]. Here we demonstrate, that one can build a superconducting SET (Cooper pair transistor) without any tunnel junctions. Instead a pair of thin superconducting titanium wires in QPS regime - the quantum phase slip junctions (QPSJ) - can be used (Fig. 1b).

## Methods

The samples were fabricated using PMMA/MAA double layer lift-off e-beam lithography and directional ultra-high vacuum metal deposition. The all-titanium nanostructures were formed on surface of oxidized silicon (Fig. 1c). Low energy directional ion milling [14] was used to reduce the cross section of the nanowire down to sub-30 nm scales. Scanning electron and atomic force microscope analyses were used to test the samples. Only those structures which contained no obvious artifacts were further processed. Electron transport measurements were made at ultra-low temperatures in $^3$He$^4$He dilution refrigerators located inside electromagnetically shielded room. All input/output lines contained multi-stage RLC filters to reduce the impact of noisy EM environment [15].

## Theory

It has been shown [16] that the Hamiltonians describing a Josephson junction (JJ)

$$\hat{H}_{JJ} = E_C \hat{q}^2 - E_J \cos(\hat{\varphi}) + \hat{H}_{COUP} + \hat{H}_{ENV} \qquad (1)$$

and a short superconducting nanowire in the regime of QPS, which correspondingly can be called the quantum phase slip junction (QPSJ)

$$\hat{H}_{QPSJ} = \frac{E_L}{(2\pi)^2}\hat{\varphi}^2 - E_{QPS}\cos(2\pi\hat{q}) + \hat{H}_{COUP} + \hat{H}_{ENV} \qquad (2)$$

are identical with accuracy of substitution $E_C \leftrightarrow E_L$, $E_J \leftrightarrow E_{QPS}$ and $\varphi \leftrightarrow \pi q/2e$, where $E_C$, $E_L$, $E_J$, $E_{QPS}$ are the energies associated with charge, inductance, Josephson and QPS couplings, $q$ is the quasicharge and *2e* is the charge of a Cooper pair. $\hat{H}_{COUP}$ and $\hat{H}_{ENV}$ are the coupling and environmental Hamiltonians, which can be similar for a JJ and a QPSJ. The identity of the Hamiltonians (1) and (2) reflects the fundamental quantum duality of these two systems.

The rate of QPSs $\Gamma_{QPS}$ is determined by the nanowire dimensions and its material parameters [5,17,18]:

$$E_{QPS} \equiv \Gamma_{QPS} h = \Delta \frac{R_Q}{R_N}\left(\frac{L}{\xi}\right)^2 \exp(-S_{QPS}) \qquad (3)$$

$\Delta$ is the superconductor energy gap, $\xi$ is the coherence length, *L* is the nanowire length. The QPS action is $S_{QPS}=A[R_Q/R_N][L/\xi(T)]$, where $R_N$ is the sample resistance in normal state, $R_Q=h/(2e)^2=6.45$ k$\Omega$ is the superconducting quantum resistance, the constant $A\approx 1$ is the numerical factor that unfortunately, cannot be determined more precisely within the model. It can be easily shown that for given (small) cross section of a nanowire, materials with low critical temperature and high normal state resistivity are of advantage for observation of the QPS effect [5].

## Results and discussion

The I-V characteristics of structures similar to the ones from Fig. 1b,c were measured at temperatures above and below the critical temperature of thin film superconducting titanium $T_c\approx 500$ mK. Above the critical temperature no non-linearity of the I-Vs was detected, while at $T<<T_c$ a pronounced Coulomb blockade was observed. The effect can be seen while recording the I-V characteristics. Measuring such dependencies is rather challenging because close to zero current bias the resistance of the system tends to infinity, signal becomes extremely noisy and additionally the corresponding RC constants become very large. To improve the signal-to-noise ratio we used modulation technique recording the dV(dI) dependencies (Fig. 2). One can clearly see that close to zero bias I→0 the dynamic resistance dV/dI indeed tends to infinity.

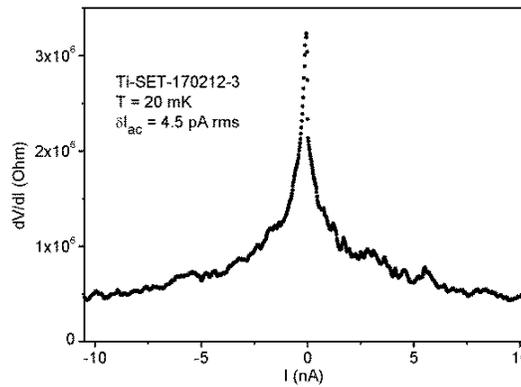

*Fig. 2. **I-V characteristics.** The dV/dI dependence of a QPSJ Cooper pair transistor measured at T=20 mK using modulation technique.*

The application of gate potential $V_{gate}$ modulates the Coulomb gap $V_c$ (Fig. 3). The period of the modulation is in a reasonable agreement with sample geometry (gate capacitance). The depth of the modulation depends on the bias point and is maximal at I→0. Application of sufficiently strong magnetic field suppresses the Coulomb blockade.

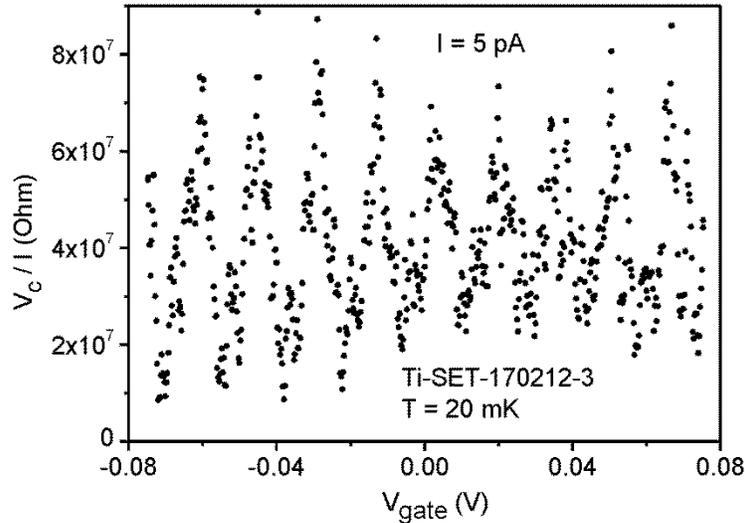

*Fig. 3. **Gate modulation**. Coulomb gap $V_c$ vs. gate potential $V_{gate}$ at a finite current bias I= 5 pA measured at temperature T=20 mK.*

The observation that the Coulomb blockade disappears above the critical temperature $T_c$ and critical magnetic field $H_c$ is very important for interpretation of the results in terms of QPSs. If at low temperatures $T<<T_c$ above the critical magnetic field a finite Coulomb gap could be observed, then one might argue that the effect originates not from QPS, but is rather related to unintentionally formed tunnel (in superconducting state - Josephson) junctions. For the known cross section of our nanowires a simple estimation gives the charging energy $E_c/k_B$ of such a hypothetical junction of about 7 K. Hence, if the observed Coulomb blockade is related to overlooked tunnel junction(s), then in the normal state ($T<<T_c≈500$ mK, but $H>H_c≈4$ T) one should observe a pronounced Coulomb gap of about $E_c/e≈600$ μV, which is not the case. Hence, we can state that the Coulomb blockade in our samples is indeed related to essentially superconducting property - the QPS, which manifests itself as a dynamic equivalent of a conventional Josephson effect with static in space and time weak links.

It is worse to note that our gap modulation dependencies $V_c(V_{gate})$ often contain single electron component. The origin of such 'quasiparticle poisoning' in pure superconducting system is not clear. Presumably, the effect may originate from presence of non-equilibrium quasiparticles generated during each QPS event [15,19].

## *Conclusions*

Junctionless all-titanium Cooper pair transistors were fabricated, where the conventional tunnel (Josephson) junctions were substituted with thin nanowires governed by quantum fluctuations of the order parameter. At ultra-low temperatures $T<<T_c$ the pronounced Coulomb blockade was observed at I-V dependencies. The magnitude of the Coulomb gap can be modulated by the gate potential. The Coulomb blockade disappears above the critical temperature, and at low temperatures can be suppressed by strong magnetic field. We interpret the results as the QPS events provide the dynamic equivalent of tunneling through a conventional Josephson junction. The effect originates from the fundamental quantum duality of these two systems.

*Acknowledgements*

The research carried out in 2015-2016 by K. Yu. Arutyunov was supported within the framework of the Academic Fund Program at the National Research University Higher School of Economics (HSE) in 2015- 2016 (grant No.15-01-0153) and supported within the framework of a subsidy granted to the HSE by the Government of the Russian Federation for the implementation of the Global Competitiveness Program. The support of COST Action MP-1201 (Nano-SC) is acknowledged.